\documentclass[aps,prc,tightenlines,floatfix,showpacs,twocolumn]{revtex4-2}
\usepackage{graphics}
\usepackage{bm}
\usepackage{amsmath}
\usepackage{amssymb}
\usepackage{txfonts} 
\usepackage{supertabular}
\usepackage{array} 
\usepackage{hyperref}
\hypersetup{colorlinks=true,citecolor=blue,urlcolor=blue,linkcolor=blue}

\newcommand{\sfrac}[2]{\textstyle\frac{#1}{#2}}

\usepackage{calrsfs}
\DeclareMathAlphabet{\pazocal}{OMS}{zplm}{m}{n}

\begin{document}

\title{Low-lying states of $^{115}${\rm Cd} modeled as a valence neutron coupled to $^{114}${\rm Cd}}
\author{K. Amos$^{1,2}$}
\altaffiliation{Deceased.}
\author{P. R. Fraser$^{3,1,4}$}
\email{paul.fraser@unsw.edu.au}
\author{S. Karataglidis$^{2,1}$}
\email{stevenka@uj.ac.za}
\author{D.~Bhattacharyya$^{3,4}$}
\email{deeparna.bhattacharyya@unsw.edu.au}
\author{L. Canton$^{4}$}
\email{luciano.canton@pd.infn.it}
\author{K. Murulane$^{2}$}
\email{kmurulane@uj.ac.za}

\affiliation{
$^{1}$ School of Physics, University of Melbourne, Victoria 3010, Australia\\
$^{2}$ Department of Physics, University of Johannesburg, P.O. Box 524,
Auckland Park, 2006, South Africa\\
$^{3}$ School of Science, The University of New South Wales, Canberra, ACT 2600, Australia\\
$^{4}$ Department of Fundamental and Theoretical Physics, The Australian National University, Acton, ACT 2601, Australia\\
$^{6}$ Istituto Nazionale di Fisica Nucleare, Sezione di Padova, I-35131, Italy\\
}

\date{\today}

\begin{abstract}
$^{114}$Cd has historically been regarded as an ideal quadrupole vibrator due to its near-canonical level spacings, though this interpretation is disputed due to electromagnetic transition data from the two- and three-phonon states. In this paper we investigate whether, with the extension of including the single-phonon octupole state, the model is still useful in describing the coupling of a neutron to a cadmium-114 core to model the lowest-energy cadmium-115 states, as the triplet states have small to no influence in the calculation. The results thus provide supporting evidence for the tentatively-assigned $J^\pi$ values of the lowest-energy $^{115}$Cd states. A more-rigorous derivation than often used for the vibrational $B(E\lambda)$ is used, in conjunction with the $^{115}$Cd structure calculation, to provide a theoretical prediction of the $^{114}$Cd octupole deformation.
\end{abstract}

\maketitle

\section{Introduction}

\subsection{Perspectives on $^{114}$\text{Cd}}

Surprisingly little theoretical investigation has been made into the structure of $^{115}$Cd, given the obvious interpretation of a valence neutron coupled to a collectively-behaving $^{114}$Cd core~\cite{Be69}. The surprise comes given that the long history of interpretation of $^{114}$Cd as an exemplary quadrupole-deformed vibrator has recently been reviewed in detail and called into question~\cite{Ga10,Ga14}, along with the applicability of the model more generally~\cite{Sh19}.

In the textbook~\cite{Bo75,Wo90,Ro10} model of a spherical nucleus (even-even and closed-shell) undergoing harmonic vibration by excitation with quadrupole phonons, the vacuum $0^+$ ground state is excited by a single $\hbar\omega$ phonon to a $2+$ state. A second phonon of equal energy will couple to the first to form a degenerate triplet consisting of a $0^+$, a $2^+$, and a $4^+$ state (with a $1^+$ and $3^+$ disappearing by symmetry arguments), and a third phonon of equal energy excites a degenerate quintuplet with members $0^+$, $2^+$, $3^+$, $4^+$, and $6^+$ (with a $1^+$ and $5^+$ also disappearing by symmetry). The reduced transition probability from an $n$-phonon state to an $n-1$ state will be proportional to $n$, so each member of the triplet of states has twice the $B(E2)$ strength to the one-phonon $2^+$ as the latter has to the ground state. Two phonon transitions will be forbidden under the simplest models. The ratio of triplet state energies to the one-phonon $2^+$ energy being 2, and sometimes more specifically $E(4^+_1)/E(2^+_1)=2$~\cite{Ke95}, has become a shorthand for likely quadrupole vibrator behaviour since early energy and $E2$ systematics supported a collective interpretation of many even-even nuclei across the wide range $36\le N\le 88$, including $^{114}$Cd~\cite{Sc55}.  

Acceptance of the quadrupole vibrator interpretation of $^{114}$Cd was enhanced when a near-degenerate triplet was soon-after identified by Coulomb excitation~\cite{McG65}, though a subsequent measurement to refine information about the $2^+_1$ state found that the $B(E2; 2^+_2\rightarrow2^+_1)/B(E2; 2^+_1\rightarrow0^+_1)$ was only around 0.7, not 2~\cite{Mi69}.
Current evaluated data lists $E(2^+_1)=558.456(2)$~keV and the triplet $0^+_2,2^+_2,4^+_1$ members at 1134.532(6)~keV, 1209.708(6)~keV, and 1283.739(8)~keV, respectively, spanning approximately 149~keV~\cite{Bl12}. Slightly higher than three times the $2^+_1$ state energy is a cluster of states spanning $\sim$270~keV that includes, with a couple of others, a $0^+$, at least one $2^+$, a $3^+$, at least one $4^+$, and a $6^+$, i.e., a set suggestive of a degenerate three-quadrupole-phonon quintuplet. Within that range is a $3^-$ state which is suggestive of a single-octupole-phonon state within the same model.

Electromagnetic transition rates, however, were measured much later than the properties of the states, which had by then established wide acceptance of the vibrator picture.  
The adopted strengths $B(E2;4^+_1~\rightarrow~2^+_1)~=~62(4)$~W.u. and $B(E2;2^+_1~\rightarrow~0^+_1)~=~31.1(19)$~W.u. give a ratio of 1.99(18), consistent with the harmonic value of 2 when their uncertainties are treated as independent. However, the $0^+_2\rightarrow2^+_1$ and $2^+_2\rightarrow2^+_1$ strengths, 27.4(17) and 22(6)~W.u., are below the corresponding harmonic expectation. The quadrupole moment $Q(2^+_1)=-0.35(5)~e\,\mathrm{b}$~\cite{Bl12} also differs from the zero moment predicted by the harmonic model assuming a spherical ground state. Finally, the $\gamma$-transitions between the quintuplet $0^+$ state and triplet $2^+$ state, and that between the quintuplet $2^+$ state and triplet $4^+$ state have not been observed~\cite{Bl12}.

Extensions beyond the harmonic description of the vibrational model, which may be collectively termed anharmonic models, rely on either coupling to the shell model \cite{Ra59} or using perturbation theory to give terms beyond the ideal vibrator \cite{Be69}. In a shell model framework, one normally assumes a space of valence nucleons outside an inert core. Raz \cite{Ra59} extends this beyond the standard shell model by assuming that the core may be described as a vibrator, with the vibrations coupling to the orbits of the valence nucleons to describe the nucleus in question. B\`es and Dussel \cite{Be69} use perturbation theory to introduce deformation into
the vibrator, leading to such additional terms. Concomitantly, the MCAS description falls partway between the two: it allows for deformation (see below), and the coupling of the projectile to the vibrating core is akin to the Raz model.

One of the appeals of interpreting $^{114}$Cd as a quadrupole vibrator is that there are few other states within the 3$\hbar\omega$ energy range.
Intruder $0^+_3$ (1.306~MeV) and $2^+_3$ (1.364~MeV) states, being at energies slightly higher than the triplet, complicate the assignment, and have produced much discussion since their discovery~\cite{Co60}. Shape co-existence has been a common interpretation of these states (and a significant review on the wider topic has been recently published~\cite{Bo23}); Ref.~\cite{Me77} proposed that the $0^+_2$ state is the head of a quasi-rotor band, with the $2^+_3$ intruder as the next member and the $4^+_2$ as the next after that. In contrast, Ref.~\cite{Ju80} used the $B(E2; 0^+_3\rightarrow2^+_2)$ (127(16)~W.u. by current measurement~\cite{Bl12}) and $B(E2; 0^+_3\rightarrow2^+_1)$ (0.0026(4)~W.u. in current measurement) to propose that the $0^+_3$ is the three-phonon $0^+$. Ref.~\cite{Sc82b} described a model with an interference term for vibration and rotation states, and in Ref.~\cite{He82} many of the same authors investigated a model wherein two-particle, two-hole excitations interact with quadrupole vibration states. Ref.~\cite{Ba86} found ``[n]o evidence for a sizable admixture of the intruder [$0^+_3$ state] in the ground state", with their cross section measurements in ``excellent agreement" with DWBA calculations. In response, Ref.~\cite{Fo87} claimed that that experiment ``did not succeed" in observing what the latter had predicted using a simplistic interacting boson approximation method that relied on many assumptions. Such comments are put into context elsewhere~\cite{Pe12,Am13,Ga13}. In any case, evaluated data~\cite{Bl12} show that the $0^+_3$ state couples strongly to the triplet $2^+_2$, with $B(E2\downarrow)=127(16)$~ W.u., and weakly to the $2^+_1$, with $B(E2\downarrow)=0.0026(4)$~ W.u., and $I_{\gamma +ce}$ of 1.4(7) and 20.8(11) for the $0^+_2$ and $0^+_1$, respectively. The $2^+_3$ has $B(E2\downarrow)$ to the $4^+_1$, $2^+_2$, $0^+_2$ and $2^+_1$ of 45(7), 27(4), 65(9) and $<$0.045~W.u., respectively. Ref.~\cite{Fa88} surveyed data to support either vibrational structure, or vibrator/rotor shape co-existence, and found the data to favour a mainly vibration-like structure.

\begin{figure}
\centering
\scalebox{1.3}{\includegraphics*{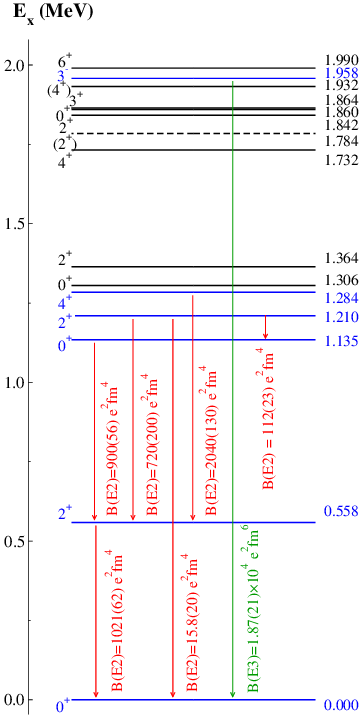}}
\caption{\label{fig:Cd114-spectrum}Observed low-energy spectrum of $^{114}$Cd from the adopted $A=114$ evaluation~\cite{Bl12}. The states used in the calculation of Section~\ref{discuss} are shown in blue. Only $B(E2)$ and $B(E3)$ values observed between them are shown. Parentheses denote tentative spin-parity assignments.}
\end{figure}

Approaches other than a quadrupole vibrator or the same plus a rotor band have been attempted. A microscopic calculation~\cite{Ko03} generated one-phonon states from a single-particle basis and a many-body Hamiltonian, with anharmonic vibrations and deformed intruder states, using the Quasiparticle Random-Phase Approximation.
They report that interpreting the $0^+_2$, $2^+_2$ and $4^+_1$ states as anharmonic two-phonon-like states gave the best $B(E2; J^+_i\rightarrow J^+_1)/B(E2; 2^+_1\rightarrow0^+_1)$, but gave few details about the alternatives explored.
Ref.~\cite{Co15} developed an effective field theory for quadrupole vibrations with quantified theoretical uncertainties and a breakdown scale near the three-phonon level. In fitting the $^{114}$Cd spectrum, the authors treated the $0^+_2$ and $2^+_2$ levels as members of the two-phonon triplet. The calculated low-lying spectrum and most electromagnetic observables were consistent with the data within the estimated uncertainties, with the diagonal $2^+_2$ quadrupole matrix element as an exception. On this basis, the authors suggested that $^{114}$Cd could be described as an anharmonic quadrupole vibrator.

A current movement in the literature proposes abandoning the vibrational interpretation in favour of a rotor picture~\cite{Ga14}, with the suggestion that the alignment between the spectral form predicted by the vibration model and the even Cd isotopes is coincidental. If the states of several rotor bands align with the expected vibrator pattern, this may explain unobserved $\gamma$-transitions, that should be in a vibrator, as being weak couplings between different rotational bands~\cite{Ga10}. The tiny value of the observed $B(E2;0^+_2\rightarrow2^+_1)$ and enhanced $B(E2;0^+_3\rightarrow2^+_2$ suggested, more specifically, $\gamma$-soft rotation~\cite{Wi56} to those authors~\cite{Ga14}. 
The Bohr Hamiltonian calculation of Ref.~\cite{Pr12} obtained a deformed minimum with weak $\gamma$ dependence for $^{114}$Cd, supporting this interpretation. 
Interpreting the $2^+_2$ as the second band head, their calculated $B(E2)$ values (ground-state band $2^+_1\rightarrow0^+_1$, $4^+_1\rightarrow2^+_1$, and $4^+_2\rightarrow2^+_2$, and inter-band $2^+_2\rightarrow0^+_1$) generally agreed with the measured trends but tended to be systematically larger. 
The largest energy discrepancies was noticed in the $0^+_3$ states, 
while the possible intruder character of the low-lying $0^+$ states remained outside the scope of that study.
In a fuller survey, Ref.~\cite{Ga18} concluded that ``the detailed studies performed over the past several decades have all but eliminated [multiphonon structures in spherical nuclei] as a structural paradigm." This idea was then further examined in Ref.~\cite{Sh19}, which compared $\gamma$ vibration and triaxial rotor state interpretations. As part of a wider review, Ref.~\cite{Le24} presents a survey of potential energy surfaces of $\beta$ and $\gamma$ parameterisations from different calculations. 

In a different approach, Ref.~\cite{Ra22} used a triaxial projected shell model to generate yrast, $\gamma-$ and $\gamma\gamma-$bands for even Cd isotopes from 104 to 122, though this spanned a larger energy range than considered here, and did not explain the full low-lying spectrum.

\subsection{Bound-state spectrum of $^{115}$Cd}

As $^{114}$Cd exhibits such a degree of collectivity, even if its kind is disputed, a logical hypothesis for the behaviour of $^{115}$Cd is a valence neutron coupled to the collective core.
The adopted spectrum of $^{115}$Cd begins with a $1/2^+$ ground state and a $(11/2)^-$ isomer at 181.0(5)~keV. The rest of the adopted spectrum is not as well-established; the present calculation of the low-lying states may shed more light on those states.

The spectrum of $^{115}$Cd may be considered in the frameworks of Raz \cite{Ra59} and
B\`es and Dussel \cite{Be69} insofar that the coupling of the projectile (valence neutron) to the $^{114}$Cd core is encompassed. The Raz approach treats the projectile as a valence shell-model neutron to couple to the vibrating $^{114}$Cd core while the B\`es and Dussel model treats the spectrum as perturbations of the pure $^{114}$Cd vibrator stemming from the coupling of the additional neutron.

These studies motivate testing how well coupling a valence neutron to a vibrational representation of selected $^{114}$Cd states can describe the low-lying $^{115}$Cd spectrum. We do this here, using the multi-channel algebraic scattering (MCAS) technique~\cite{Am03,Ka19} as a structure approach. MCAS is a momentum-space formalism wherein negative projectile energies give bound states of the projectile and core. This approach has the advantage of accounting for the Pauli principle in coupling the nucleon to multiple states of a collective core. In the following section, we summarise the model and its use in defining the matrix of potentials for the coupled-channel problem. Results of using this approach are given in Section~\ref{discuss}. In Section~\ref{em}, we survey electromagnetic transition rates and how they are informed by and inform deformation parameters as well as charge distributions. Finally, conclusions are drawn in Section~\ref{conc}. 


\section{The MCAS method}
\label{MCAS}

In order to study the coupling of a valence neutron to $^{114}$Cd we use the Multichannel Algebraic Scattering (MCAS) method~\cite{Ka19,Am03}. Despite being a scattering formalism, it functions as a structure calculation method, as being in momentum space allows negative-energy scattering to find bound-state resonances corresponding to compound system eigenstates. It does this by solving coupled-channels Lippmann-Schwinger equations:

\begin{align}
T_{cc'}(p,q;E) &= V_{cc'}(p,q) \nonumber\\
&+ 
\ \frac{2\mu}{\hbar^2}
\left[ \sum_{c'' = 1}^{\rm open} \int_0^\infty 
\frac{V_{cc''}(p,x) \ \ T_{c''c'}(x,q;E)}
{k^2_{c''} - x^2 + i\epsilon}\ x^2\  
dx \right.
\nonumber\\
&\, \left.\ \ - \sum_{c'' = 1}^{\rm closed} \int_0^\infty 
\frac{V_{cc''}(p,x) \ \ T_{c''c'}(x,q;E)}
{h^2_{c''} + x^2} 
\ x^2\ dx \right], 
\label{LSeqn}
\end{align} 
where $\mu$ is the reduced mass of target and projectile, and the $c$ are indices denoting the individual channels. Open and closed channels have wave numbers $k_c$ and $h_c$, respectively. The $V_{cc'}$ are interaction potentials between the valence neutron and core. These directly incorporate nuclear structure information of the core, and while in principle any structure model may be used, a Tamura collective model of rotor or vibrator character covers the primary bands often observed at low energies~\cite{Ta65}.

To solve the Lippmann-Schwinger equations, the potentials are expanded into a finite number of energy-independent, separable terms called sturmians, $\chi$~\cite{We65}, as outlined in Ref.~\cite{Am03}. The $S$-matrix is defined in terms of these sturmians as: 
\begin{multline}
  S_{cc'} = \delta_{cc'} \;-\; i^{l_{c'} - l_c + 1} \pi \mu \\ \times\sum^{N}_{n,n'
  = 1} \sqrt{k_c} \hat{\chi}_{cn}(k_c) \left( \left[ {\mbox{\boldmath
  $\eta$}} - \mathbf{G}_0 \right]^{-1} \right)_{nn'} 
  \hat{\chi}_{c'n'}(k_{c'}) \sqrt{k_{c'}}
\end{multline}
where
\begin{multline}
  \left[ \mathbf{G}_0 \right]_{nn'} = \mu \left[
  \sum^{\text{open}}_{c = 1} \int^{\infty}_0 \hat{\chi}_{cn}(x)
  \frac{x^2}{ k^2_c - x^2 + i\epsilon } \hat{\chi}_{cn'}(x) \, dx \right. \\ - 
   \left. \sum^{\text{closed}}_{c = 1}
  \int^{\infty}_{0} \hat{\chi}_{cn}(x) \frac{ x^2 }{ h_c^2 + x^2 }
  {\hat{\chi}_{cn'}(x)} \, dx \right] \ ,
\label{Gmatrix}
\end{multline}
with this form of Green's function being suitable for particle-stable nuclear target states. An alternate form for particle-unstable targets is outlined in Refs.~\cite{Fr08a,Fr16a}.

\subsection{Vibrational-model scattering potentials}
\label{vib-model}

The basic form of the interaction potential is
\begin{equation}
V(r) = 
\left[ V_0 + V_{ll} {\{\bf l \cdot l}\} + V_{ss} {\{\bf I \cdot s}\} 
\right] f(r) + 2\lambda_\pi^2 
V_{ls} \frac{1}{r} \frac{\partial f(r)}{\partial r} {\{\bf l \cdot s}\}\ ,
\label{poteq}
\end{equation}
where $f(r)$ is the Woods-Saxon (WS) function,
\begin{equation}
f(r) = \left[1 + \exp\left(\frac{r - R_0}{a_0} \right) \right]^{-1} \ . 
\end{equation}
$V_0$, $V_{ll}$, $V_{ss}$ and $V_{ls}$ are thus the parameters of the model. 

Modifying this base potential to incorporate surface deformation, $\varepsilon$, serves to include the structure model information \textit{viz.}
\begin{equation}
R(\theta \phi) = R_0 \left[1 + \varepsilon \right] \ = R_0 \left[ 1 +
\sum_{\lambda > 1, \mu} \alpha_{\lambda \mu}^\star
Y_{\lambda \mu}(\theta \phi) \right] \ ,
\label{dropsur}
\end{equation}
where $R$ is the deformed radius, $R_0$ is the spherical radius, and the perturbations, $\alpha$, are built from phonon creation and annihilation operators, $b^\dagger_{\lambda \mu}$ and $b_{\lambda \mu}$:
\begin{equation}
\alpha_{\lambda \mu} \Rightarrow \sqrt{\frac{\hbar}{2B_\lambda \omega_\lambda}}
\left[b_{\lambda \mu} + (-)^\mu b^\dagger_{\lambda -\mu} \right]\ .
\end{equation}
Taking the expansion to second order,
\begin{equation}
f(r) = f_0(r) + \varepsilon \left(\frac{\partial f(r)}{\partial \varepsilon} \right)_0 
+ \frac{1}{2} \varepsilon^2 \left(\frac{\partial^2f(r)}{\partial \varepsilon^2} \right)_0\ . 
\end{equation}
and treating $R(\theta, \phi)$ as the variable in \linebreak $f(r) = f(r-R(\theta, \phi))$, the potential may be cast in terms of derivatives of the nuclear radius, \textit{viz.}
\begin{multline}
f(r) = f_0(r) - R_0 \sum_{\lambda \mu} \alpha_{\lambda \mu}^\star
Y_{\lambda \mu}(\theta, \phi) \left(\frac{\partial f(r)}{\partial r}\right)_0\\ 
+ \frac{1}{2} 
R_0^2 \sum_{l_1 m_1 l_2 m_2} \alpha_{l_1 m_1}^\star \alpha_{l_2 m_2}^\star
Y_{l_1 m_1}(\theta, \phi) Y_{l_2 m_2}(\theta, \phi) \left(\frac{\partial^2f(r)}{\partial r^2} 
\right)_0\ ,
\label{Vib-fst}
\end{multline}
with full details in Ref.~\cite{Sv17}. A similar expansion is made for the spin-orbit contribution, $g(r) = \frac{1}{r}\frac{\partial f(r)}{\partial r}$. 

In this calculation, core nucleus states are constructed from the vacuum as one quadrupole or octupole phonon excitations, or combinations of two of those phonons, leading to a set of 13 natural spin-parity configurations. We restrict to natural $J^\pi$ states as these couple more strongly. Further, couplings are also restricted to differences of two phonons at most. Hence, states with one quadrupole (octupole) phonon are not coupled to those with two octupole (quadrupole) phonons, as these differ by 3 phonons. Nor are states from two quadrupole and two octupole phonons coupled. Thus, of the 13 configurations, we restrict the target state set to 10: two $0^+$ states, two $2^+$, two $3^-$, and one state each with $J^{\pi} = 1^-$, $4^+$, $5^-$ and 6$^+$. This accommodates the low-energy spectrum of most light mass nuclei that may be described by a vibrational model.

\subsection{Treatment of the Pauli principle}

In nucleon-nucleus (NA) scattering, the nuclear target is commonly treated as a collective object that reproduces the relevant quantum properties of the system without explicit reference to its nucleon constituents. While computationally convenient, this approximation neglects antisymmetrization constraints imposed by the Pauli exclusion principle on the interacting nucleons.

In single-channel calculations, Pauli compliance is achieved naturally: deep-lying forbidden states in the single-particle potential are orthogonal to all physical eigenstates by construction, and the scattering wave function inherits this orthogonality automatically.

In coupled-channel (CC) frameworks, however, a more subtle difficulty arises. Any actual compound state of a definite $J^\pi$ is formed as a superposition of all uncoupled basis states sharing that $J^\pi$, including any that are Pauli-forbidden. The resulting coupled-channels wave functions can therefore acquire spurious components. 

The Pauli principle can be viewed as a restriction on the region of Hilbert space that the system is otherwise allowed to occupy. Manually excluding these regions is effectively the same as enforcing the Pauli principle. In practice, this is achieved by altering the matrix of input coupled-channel scattering potentials to incorporate channel-dependent auxiliary potentials, known as orthogonalising pseudo-potentials (OPP)~\cite{Ku78,Kr74}, a generalization of the orthogonality condition model of Saito~\cite{Sa69}. When used to construct Sturmian basis functions, the modified potential takes the form

\begin{equation}
V_{cc'}(r,r') = V_{cc'}(r)\,\delta(r-r') 
+ \lambda_c\, A_c(r)\, A_c(r')\, \delta_{c,c'}\ ,
\label{OPPeq}
\end{equation}

where $V_{cc'}(r)\,\delta(r-r')$ is the local collective-model potential, and the supplementary terms $\lambda_c A_c(r) A_c(r') \delta_{c,c'}$ are the OPP: nonlocal in coordinate space and diagonal in channel index. The function $A_c(r)$ spans the excluded phase space in channel $c$, obtained as the bound-state solution of the radial Schr\"{o}dinger equation,

\begin{equation}
\frac{d^2}{dr^2}A_c(r) + \left\{
\frac{2m}{\hbar^2}\left[E - V_{cc}(r)\right] 
- \frac{l(l+1)}{r^2}
\right\} A_c(r) = 0\ .
\label{Ac-eq}
\end{equation}

The blocking strength $\lambda_c$ is, in principle, infinite for a completely forbidden state; in practice, $\lambda_c = 1$~GeV is sufficient to enforce orthogonality in the Sturmian construction, ensuring the 
separable expansion of the interaction inherits the same property.

For orbitals that are only partially occupied in the microscopic nuclear description, a reduced value of $\lambda_c$, referred to as \textit{Pauli hindrance} may be employed. Tuning $\lambda_c$ in this way provides a means of adjusting the resonance structure of the scattering cross section to better match the spectrum of the compound nuclear system. The OPP are applied exclusively to channels $c$ whose quantum numbers correspond to occupied shell-model configurations. The validity of this treatment is supported by the consistency in the analysis, within the MCAS framework, comparing predictions and the observed spectrum of $^{13}$C, when a neutron couples to the $^{12}$C target \cite{Am03,Ca05}.

\section{Results and discussion}
\label{discuss}

The spectrum of $^{114}$Cd, showing the target states used in the MCAS calculations to evaluate the states in $^{115}$Cd, is shown in Fig.~\ref{fig:Cd114-spectrum}, together with the electromagnetic transition probabilities between select states. Of the $E2$ transitions shown, those involving the $2^+_1$ state are strong. Only the $2^+_2 \rightarrow 0^+_1$ and the $2^+_2 \rightarrow 0^+_2$ transitions are weak. We do not include the latter transition given that it is within the second cluster of states and does not involve any phonon exchange. The $E3$ transition to the ground state shown in Fig.~\ref{fig:Cd114-spectrum} is also strong and is included in the MCAS calculations.

\begin{figure*}
\centering
\scalebox{0.5}{\includegraphics*{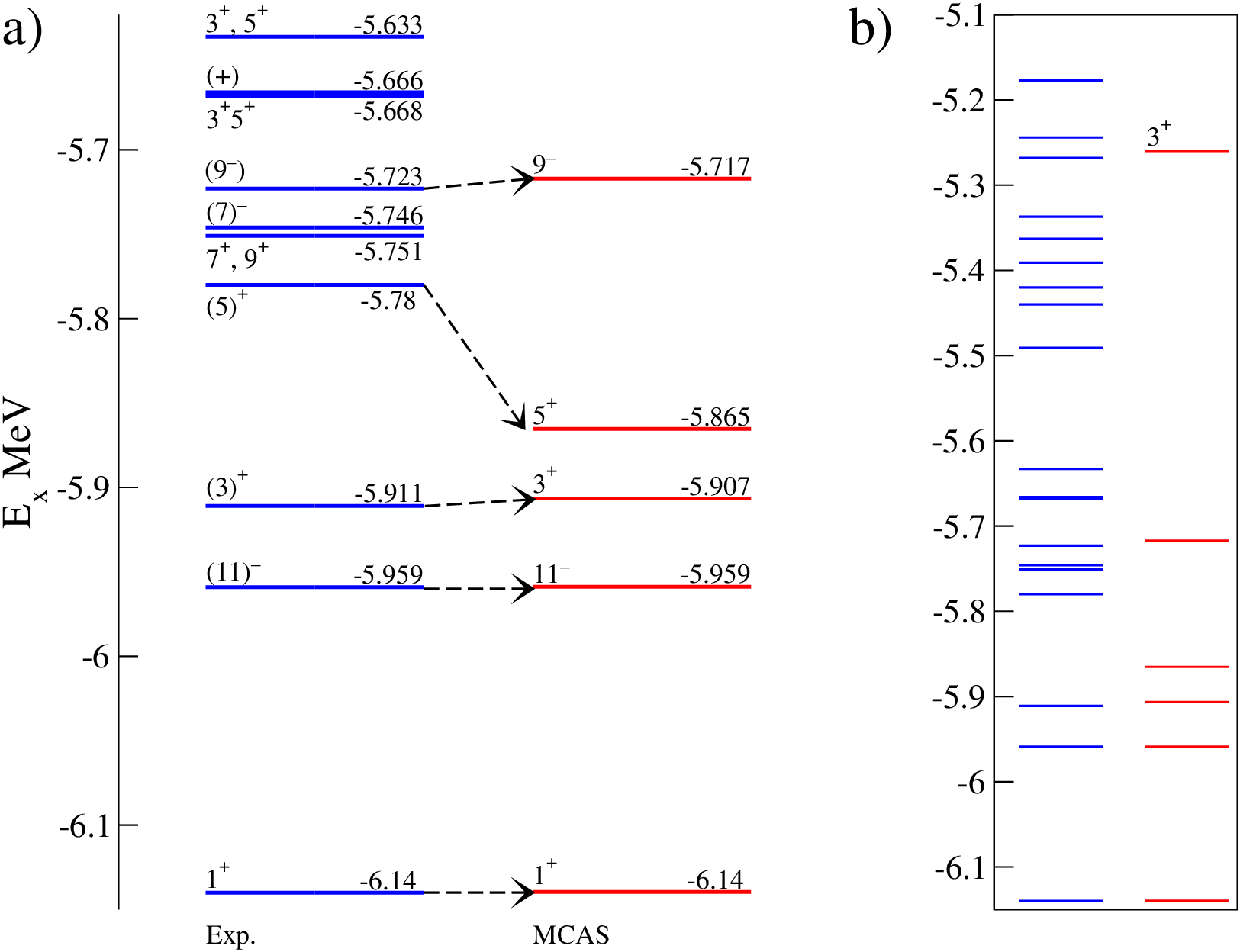}}
\caption{\label{115Cd_spec}Experimental spectrum of $^{115}$Cd~\cite{Bl12a} compared to the subthreshold bound states of a MCAS $^{114}$Cd$(n,n)^{114}$Cd calculation. Panel a) shows the lowest few states, while panel b) shows states in the first MeV. The states are classified by $2J^\pi$.}
\end{figure*}

Using the potential parameters given in Table~\ref{Cd-param} and deformation parameters as determined from the information given in Table~\ref{Elec}, we obtained the spectrum of the states in $^{115}$Cd shown in Fig.~\ref{115Cd_spec}. Note that the OPP weights are for the most part $10^6$, with small values affecting the $1d$, $2s$ levels for the ground and $2^+_1$ states.

\label{IntPot}
\begin{table}[htp]\centering
\caption
{\label{Cd-param} 
Parameter values defining the $n+^{114}$Cd interaction.}
\begin{supertabular}{>{\centering}p{28mm} p{27mm}<{\centering} 
p{27mm}<{\centering}}
\hline
\hline
 & Odd parity & Even parity\\
\hline
$V_0$ (MeV)           & -53.59 & -46.63 \\
$V_{l l}$ (MeV)       & -0.16   & -0.16 \\
$V_{l s}$ (MeV)       & 0.7 & 0.7 \\
$V_{ss}$ (MeV)        & 1.48 & 1.48 \\
\end{supertabular}
\begin{supertabular}{>{\centering}p{13mm} p{17mm}<{\centering} 
>{\centering}p{16mm} >{\centering}p{16mm} p{16mm}<{\centering} }
\hline
\hline\\[-2ex]
Geometry & $R_0 = 5.4$ fm  & $a = 0.6$ fm & $\beta_2 = 0.19$ & $\beta_3 = 0.155$ \\ 
\end{supertabular}
\begin{supertabular}{>{\centering}p{17mm} p{16mm}<{\centering} 
p{15mm}<{\centering} p{15mm}<{\centering} p{15mm}<{\centering}}
\hline
\hline
  & $0s_{1/2}$ - $0g_{7/2}$ & $1d_{5/2}$ & $1d_{3/2}$ & $2s_{1/2}$\\
\hline\\[-2ex]
 $0^+_1 \; \lambda^{(OPP)}$ & 10$^6$ & 0.0 & 0.0 & 0.0\\
 $2^+_1 \; \lambda^{(OPP)}$ & 10$^6$ & 0.0 & 0.0 & 0.0\\
 $0^+_2 \; \lambda^{(OPP)}$ & 10$^6$ & 10$^6$ & 0.0 & 0.0 \\
 $2^+_2 \; \lambda^{(OPP)}$ & 10$^6$ & 10$^6$ & 0.0 & 0.0 \\
 $4^+_1 \; \lambda^{(OPP)}$ & 10$^6$ & 10$^6$ & 0.0 & 0.0 \\
 $3^-_1 \; \lambda^{(OPP)}$ & 10$^6$ & 10$^6$ & 0.0 & 0.0 \\
\hline
\hline
\end{supertabular}
\end{table}

The ground, first and second excited states are well reproduced by the results of the MCAS calculation. This includes the $\frac{11}{2}^-$ state which, in a shell-model description, would be the coupling of the neutron to the $0h_{11/2}$ orbit. With the spin-orbit splitting of the $0h$ shell, the $0h_{11/2}$ orbit lies in the middle of the $sdg$ shell and that allows for the low-lying first excited state.

The known low-energy spectrum of $^{115}$Cd, as shown in Fig.~\ref{115Cd_spec}(b), 
portrays a large level density in the first MeV above the ground state. Yet the states obtained with this model of MCAS are few in number by comparison, indicating that a larger model beyond the quadrupole and octupole phonon excitations admitted herein is required. Even so, the level of agreement obtained for those states we have obtained illustrates the utility of the method to determine structure of the states in the compound nucleus based on the vibrational model.

\begin{figure*}
\centering
\begin{minipage}[b]{0.48\linewidth}
\begin{center}
\scalebox{0.55}{\includegraphics*{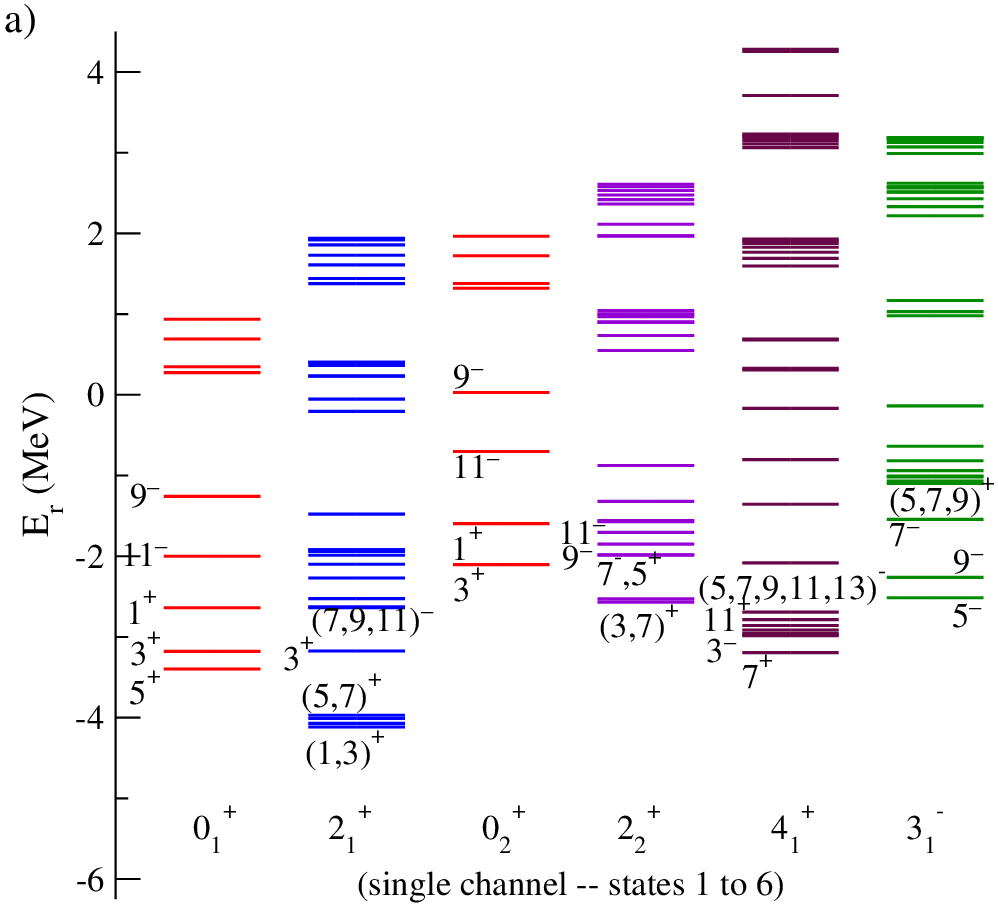}}
\end{center}
\end{minipage}\hfill
\begin{minipage}[b]{0.48\linewidth}
\begin{center}
\scalebox{0.55}{\includegraphics*{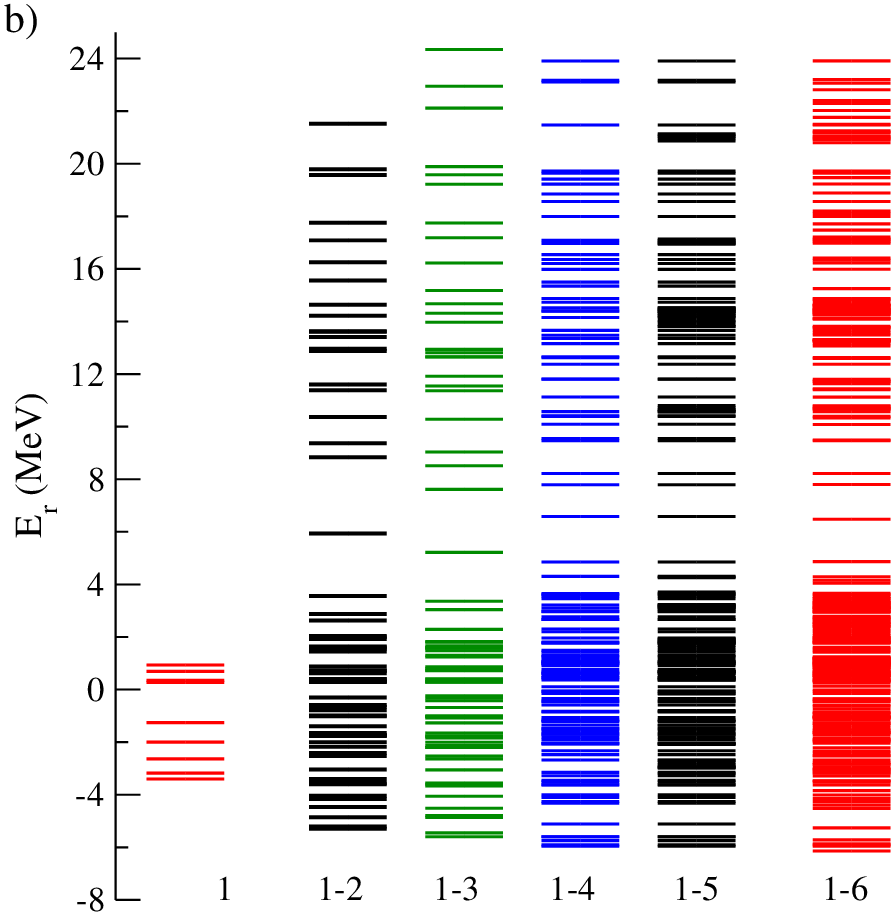}}
\end{center}
\end{minipage}
\vspace{0.7cm}

\scalebox{0.55}{\includegraphics*{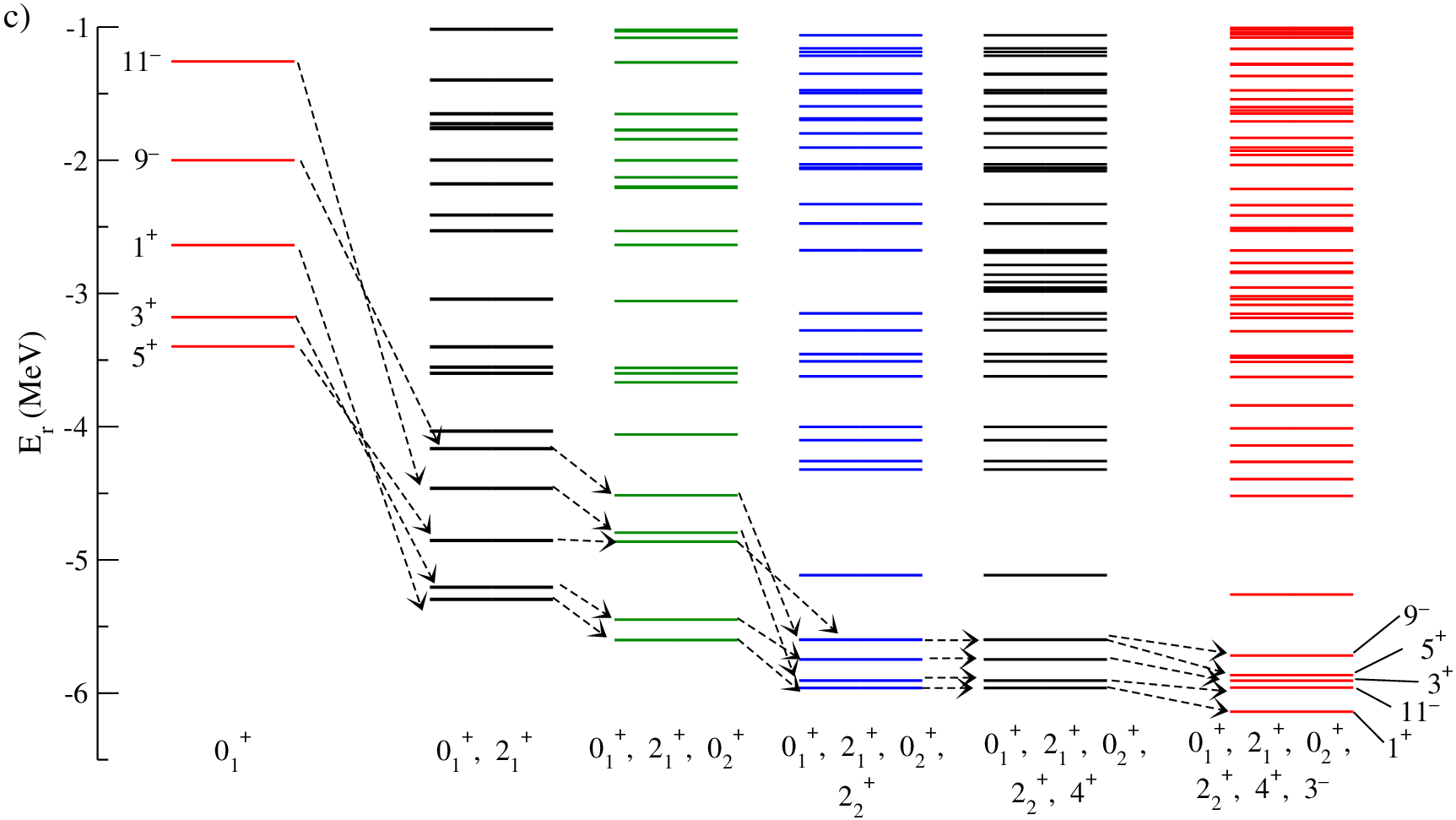}}
\caption{\label{115Cd_decomp} Low-energy bound and resonance states in $^{115}$Cd as obtained by MCAS. The energies are relative to the $^{114}$Cd$+n$ threshold. 
Panel a): Low-lying states formed when only the individual target states are used in calculations. Panel b) Changes wrought by adding channels in the calculation. 
Panel c): Detail of b). The individual states are labelled by twice their spin and their parity.}
\end{figure*}

The structure of the states in $^{115}$Cd in terms of the coupling of a neutron to vibrational states in $^{114}$Cd is further illustrated in Fig.~\ref{115Cd_decomp}.
Figure~\ref{115Cd_decomp}(a) shows the results of the calculations made using a single-channel version of the MCAS model, identifying the dominant state in $^{114}$Cd contributing to the compound states. That shows that the ground state of $^{115}$Cd, which has $J^\pi = 1/2^+$, is dominantly based on the coupling of the $2^+_1$ state in $^{114}$Cd with the neutron, with an additional component of the coupling of the ground state of $^{114}$Cd with the neutron. Figure~\ref{115Cd_decomp}(c) shows that the structure of the ground state also has components from the inclusion of quadrupole phonons involving the second cluster of states in $^{114}$Cd, but only those phonon excitations involving the $0^+_2$ and $2^+_2$ states. Couplings to the $4^+_1$ state are small by comparison and do not significantly contribute to the structure of the ground state of $^{115}$Cd.

Conversely, the first excited state in $^{115}$Cd is formed by an admixture of couplings of states of the target, notably the ground and $2^+_1$ states, with the incoming neutron as indicated in Fig.~\ref{115Cd_decomp}(a). 
Fig.~\ref{115Cd_decomp}(c) shows how the calculated $^{115}$Cd spectrum changes as target states are added successively. From left to right, the columns progress from a calculation with the target ground state alone to the full six-state calculation.
This shows that the admixtures of the couplings of the target states of the second group in the vibrational model are more significant in the structure of the $\sfrac{11}{2}^-$ state compared to the ground state. This is illustrated by the result that the $\sfrac{11}{2}^-$ state starts as the fifth excited state in the single-channel case eventually becoming the first excited state when all coupled channels are admitted. Even so, coupling involving the $4^+_1$ state of $^{114}$Cd is observed not to contribute to the structure of the $\sfrac{11}{2}^-$ state as is the case with the ground state. 

Figure~\ref{115Cd_decomp}(b) shows the spectrum of $^{115}$Cd as calculated by MCAS from a single-channel calculation to the results of the full coupled-channel calculation involving all six states of the target.
As one progresses from the single-channel calculation to the full one, the level density of the low-energy spectrum of $^{115}$Cd increases until convergence is reached, with two distinct groups of states becoming apparent. The first group extends from $-4$~MeV to 4~MeV, as shown in Fig.~\ref{115Cd_decomp}(b), and the second lies above 12~MeV. Note that there is very little difference in the spectrum as obtained with the 4-state and 5-state models. The 5-state model includes the $4^+_1$ state in $^{114}$Cd, which couples to the ground state via a two-phonon exchange. The results of
the 1-4 and 1-5 model calculations suggest that there is very little contribution from the $4^+$ state in the structure of the states in $^{115}$Cd. This supports the conclusion reached specifically for the ground and $11/2^-$ states.

The dominance of the quadrupole phonons involving only the $0^+$ and $2^+$ states of $^{114}$Cd is similar for the structures of all the low-lying bound states of $^{115}$Cd displayed in Fig.~\ref{115Cd_decomp}(c). It should also be noted that the coupling to the $3^-$ state in $^{114}$Cd from the octupole phonon also does not contribute significantly to any of the states shown in Fig.~\ref{115Cd_decomp}(c). This is a significant result and further illustrates the utility of the vibrational model used herein despite $^{114}$Cd not being a pure vibrational
nucleus.

\section{Electromagnetic transition properties}
\label{em}

As noted above, $^{114}$Cd may not necessarily be described as pure vibrator. The symmetry may be broken or there may be evidence of mixing of low-lying intruder states. To assess the utility of a vibrational model description of $^{114}$Cd, we look to the electromagnetic properties of the nucleus and evaluate the $B(E2)$ and $B(E3)$ values for the transition from the ground state to the $2^+_1$ and $3^-_1$ states, respectively. For a vibrational model of the nucleus, the forms for the relevant reduced transition probabilities, as required to generate the MCAS $NA$ coupling potential, are given in Ref.~\cite{Ka19}. We give a brief summary herein. Note that we expand the electric multipole operators to second order. This is necessitated by describing transitions between states requiring two phonons. (Expansion to first order assumes that $B(E\lambda) = 0$.)
This extension to second order also includes $B(E2)$ and $B(E3)$ for a mixed-mode description of states built from admixtures of one- or two-phonon pure vibrational basis states \cite{Am21}, but such is not needed here to assess the description of $^{114}$Cd as a pure vibrator.

The starting point for determining the $B(E\lambda)$ is the vibrational model itself and the relevant deformation parameters $\beta_\lambda$. The formalism described in Section~\ref{IntPot} uses the listed $\beta_2$ \cite{Pr16}. However, we let $\beta_3$ be a free parameter to obtain the best fit, and verify its validity in Section~\ref{em}.

One also needs a model of the charge distribution of the nucleus. While there is no Coulomb potential involved in the interaction between projectile and target, as given in Section~\ref{IntPot}, we do require a description of the charge density of the target to determine the $B(E\lambda)$ and, by extension, the relevant deformation parameters. The charge density is then constrained to reproduce the $R_\mathrm{rms}$ and $B(E\lambda)$ values.

In a vibrational description of the structure of the target nucleus, the
$B\left( E\lambda \right)$ for transitions involving one- and two-phonon exchanges is the square of the sum of the terms corresponding to the exchanges:
\begin{equation}
B\left( E\lambda \right) = \left| \mathcal{M}\left( E\lambda \right)^{(1)} + 
\mathcal{M}\left( E\lambda \right)^{(2)} \right|^2,
\label{BEtheory}
\end{equation}
where the first-order operator, for one-phonon exchange, is
\begin{equation}
    \mathcal{M}\left( E\lambda \right)^{(1)} = \frac{1}{\sqrt{2J_i+1}}
    \left\langle J_f \left\| \mathcal{C}^{(1)}_\lambda \mathcal{Q}^{(1)}_\lambda \right\|
    J_i \right\rangle
    \label{amp1}
\end{equation}
\begin{widetext}
and the second-order operator, for two-phonon exchange, is
\begin{equation}
    \mathcal{M}\left( E\lambda \right)^{(2)} = \frac{1}{\sqrt{2J_i+1}} \left\langle J_f \left\|
    \mathcal{C}^{(2)}_\lambda \sum_{l_1l_2} \sqrt{\frac{(2l_1+1)(2l_2+1)}{4\pi(2\lambda + 1)}}
    \left\langle \left. l_1 \; 0 \; l_2 \; 0 \; \right| \lambda\; 0 \right\rangle 
    \mathcal{Q}^{(2)}_\lambda \right\| J_i \right\rangle.
    \label{amp2}
\end{equation}

In Eqs.~(\ref{amp1}) and (\ref{amp2}),
\begin{equation}
    \mathcal{Q}_{\lambda\mu}^{(1)} = \beta_\lambda \frac{1}{\sqrt{2\lambda + 1}} \left\{
    b^{\dagger}_{\lambda\mu} + \left( -1 \right)^{\mu} b_{\lambda,-\mu} \right\}
\end{equation}
and
\begin{multline}
    \mathcal{Q}^{(2)}_{\lambda\mu} = \sum_{l_1l_2} \frac{1}{\sqrt{\left( 2l_1+1 \right)
    \left( 2l_2 + 1 \right)}}\beta_{l_1} \beta_{l_2} \sum_{m_1m_2} \left\langle \left.
    l_1 \; m_1 \; l_2 \; m_2 \; \right| \lambda \; \mu \right\rangle  \\
    \times \left\{ b^{\dagger}_{l_1m_1} b^{\dagger}_{l_2m_2} + \left( -1 \right)^{m_1}
    b_{l_1,-m_1} b^{\dagger}_{l_2m_2} + b^{\dagger}_{l_1m_1}\left( -1 \right)^{m_2} b_{l_2,-m_2}
    + \left( -1 \right)^{m_1 + m_2}b_{l_1,-m_1} b_{l_2,-m_2} \right\},
\end{multline}
where $b^{\dagger}_{\lambda\mu}$ and $b_{\lambda\mu}$ are the phonon creation and annihilation
operators, respectively, and the $\beta_\lambda$ are the deformation parameters.
\end{widetext}

In Eqs.~(\ref{amp1}) and (\ref{amp2}), for a general charge distribution,
\begin{align}
\mathcal{C}_\lambda^{(1)} & = - \rho_0 \, R_0\int_0^\infty r^{\lambda+2} \, \frac{\partial\rho_{\mathrm{ch}}(r)}{\partial r} dr, \nonumber \\
\mathcal{C}_\lambda^{(2)} & = \frac{1}{2} \, \rho_0 \, R_0^2 \int_0^\infty r^{\lambda+2} \, \frac{\partial^2\rho_{\mathrm{ch}}(r)}{\partial r^2} dr \ .
\label{C1C2}
\end{align}
One possible choice for $\rho_{\mathrm{ch}}(r)$ is the three-parameter Fermi (3pF) charge distribution,
\begin{equation}
    \rho_{\mathrm{ch}}(r) =  \frac{1 + w\left(\frac{r}{R_c}\right)^2}{1+\exp\left(\frac{r-R_c}{a_c}\right)} \ ,
    \label{3pF}
\end{equation}
where $\rho_0 = 4\pi\int_0^\infty r^2 \rho(r) \, dr$ $\rho_{ch}(\vec{r})$. The three parameters determining the charge distribution ($w, R_c,$ and $a_c$), together with the deformation parameters entering the $B(E\lambda)$ calculation, then provide a good representation of the Coulomb interaction between a proton and a target nucleus resulting in the spectrum of the compound system \cite{Fr16}.

The resulting $B(E\lambda)$ values of relevance in the present work are:
\begin{align}
B(E2;2^+_1 \rightarrow 0^+_1) \ &= \ \sfrac{1}{5}\, \beta_2^{\ 2} \ {\mathcal{C}_2^{(1)}}^{2}\nonumber\\
B(E2;4^+_1 \rightarrow 2^+_1) \ &= \ \sfrac{2}{9}\, \beta_2^{\ 2} \ {\mathcal{C}_2^{(1)}}^{2}\nonumber\\
B(E2;2^+_2 \rightarrow 2^+_1) \ &= \ \sfrac{2}{5}\, \beta_2^{\ 2} \ {\mathcal{C}_2^{(1)}}^{2}\nonumber\\
B(E2;0^+_2 \rightarrow 2^+_1) \ &= \ 2\, \beta_2^{\ 2} \ {\mathcal{C}_2^{(1)}}^{2}\nonumber\\
B(E2;2^+_2 \rightarrow 0^+_1) \ &= \ \sfrac{1}{35\pi}\, \beta_2^{\ 4} \ {\mathcal{C}_2^{(2)}}^{2}\nonumber \\
B(E3;3^-_1 \rightarrow 0^+_1) \ &= \ \sfrac{1}{7}\, \beta_3^{\ 2} \ {\mathcal{C}_3^{(1)}}^{2}.
\label{2ndOrd}
\end{align}
The hexadecapole transition, $4^+_1 \rightarrow 0^+_1$, is not included in the calculation of deformation as none of the states of relevance in $^{115}$Cd are formed by the coupling of the projectile to the $4^+$ of $^{114}$Cd.

A determination of the parameters for the 3pF charge distribution is found in Ref.~\cite{Vr87}, wherein the value $R_{\mathrm{rms}}~=~4.629$~fm from elastic electron scattering for $^{114}$Cd \cite{Gi75} was used to fit the parameters. 
The problem therein, however, is that $w = 0$ was used in the determination, rendering that result a two-parameter Fermi (2pF) distribution, giving $R_{\mathrm{rms}} = 4.635$~fm. Similar results were 
obtained in \cite{Li76,Fr95,Ha18,Oh22} using 2pF functional forms.

In Section 2.3 of Ref.~\cite{Ro10}, electromagnetic transition strengths are defined for a simpler vibration model assuming a uniform charge density for the nucleus and which is expanded only to first order in deformation. This yields the result (in Eq.~(2.21) of \cite{Ro10}) that $B(E2; 
0^+_2,2^+_2,4^+_1\rightarrow 2^+_1) = 2 B(E2; 2^+_1\rightarrow 0^+_1)$, (whereas the multipliers in Eq.~(\ref{2ndOrd}) are 10, 2 and $\sfrac{10}{9}$, respectively) and that $B(E2; 2^+_2,4^+_1\rightarrow 0^+_1) = 0$. This model recreates the known $B(E2; 2^+_1\rightarrow 0^+_1)$, given as 0.10~e$^2$b$^2$ where relevant. 

\begin{table*}
\centering
\caption{\label{Elec}
$R_{\mathrm{rms}}$ for $^{114}$Cd from literature and associated 2pF and 3pF charge distributions 
with the 
resulting $B(E\lambda)$ values determined in the present work. For comparison, 
values are presented from a simple model and from data. (Units have been converted from W.u. and 
$e^2$b$^\lambda$ to 
$e^2$fm$^{2\lambda}$ where appropriate.) All results from Eq.~(\ref{2ndOrd}) use $\beta_2=0.19$ and $
\beta_3=0.155$.}
\begin{supertabular}{>{\centering}p{33mm}| p{15mm}<{\centering} 
>{\centering}p{15mm} >{\centering}p{15mm} >{\centering}p{15mm} >{\centering}p{15mm} p{22mm}<{\centering} }
\hline
\hline\\[-1.9ex]
                & \cite{Gi75} & \cite{Ka71} & \cite{Fr95} & \cite{Gi75} & &\\
$R_{\mathrm{rms}}$\,(fm) & 4.629       & 4.632       & 4.610       & 4.629       & &\\
\hline
                & \cite{Vr87}$\dag$ & \cite{Li76} & \cite{Fr95} & Here & & \\
$R_c$\,(fm)     & 5.4         & 5.308       & 5.3954      & 5.536 & &\\
$a_c$\,(fm)     & 0.537       & 0.5698      & 0.52        & 0.428 & &\\
$w$             & 0           & 0           & 0           & 0.014 & &\\
\hline
                & Eq.~\ref{2ndOrd} & Eq.~\ref{2ndOrd} & Eq.~\ref{2ndOrd} & Eq.~\ref{2ndOrd} & 1$^\text{st}$-order model~\cite{Ro10} & Data \cite{Bl12}\\[3.3ex]
$B(E2;2^+_1 \rightarrow 0^+_1)$ \ ($e^2$fm$^4$) & 936  & 923  & 950  & 1022 & 1000 & 1021 \\[0.7ex]
$B(E2;4^+_1 \rightarrow 2^+_1)$ \ ($e^2$fm$^4$) & 1070 & 1026 & 1056 & 1135 & 2000 & 2036 \\[0.7ex] 
$B(E2;2^+_2 \rightarrow 2^+_1)$ \ ($e^2$fm$^4$) & 1926 & 1846 & 1901 & 2044 & 2000 & 722 \\[0.7ex]
$B(E2;0^+_2 \rightarrow 2^+_1)$ \ ($e^2$fm$^4$) & 9360 & 9232 & 9503 & 10218 & 2000 & 899 \\[0.7ex]
$B(E2;2^+_2 \rightarrow 0^+_1)$ \ ($e^2$fm$^4$) & 5.29 & 4.94 & 5.28 & 5.85 & 0    & 15.8\\
\hline
               & Eq.~(\ref{2ndOrd}) & Eq.~(\ref{2ndOrd}) & Eq.~(\ref{2ndOrd}) & Eq.~(\ref{2ndOrd}) & & Data \cite{Fe85}\\
$B(E3;3^-_1 \rightarrow 0^+_1)$ \ ($e^2$fm$^6$) & 17948 & 17198 & 17457 & 18771 & & $(1.87 \pm 
0.21)\times 10^4$\\
\hline
\hline
\end{supertabular}

$\dag$ See note in text.
\end{table*}

Table~\ref{Elec} summarises this information and the $B(E\lambda)$ values determined from Eq.~(\ref{2ndOrd}) for the 2pF parameters used in Refs.~\cite{Vr87,Li76,Fr95}. 
Results are compared to $B(E2)$ data taken from Ref.~\cite{Bl12}, which are converted from Weisskopf units to $e^2$fm$^4$ using the usual conversion~\cite{Wo90}:
\begin{equation}
    B_W(E\lambda) = \frac{1}{4\pi} \left( \frac{3}{\lambda+3} \right)^2 1.2^{2\lambda} A^{2\lambda/3}  B(E\lambda) \left( e^2 \text{fm}^{2\lambda} \right) .
    \label{Weisskopf}
\end{equation}
$B(E3;3^-_1 \rightarrow 0^+_1)$ has not been measured, but we use the measured value of $B(E3;3^-_1 \rightarrow 0^+_1)$ and the identity Eq.~(1A-68) of Ref.~\cite{Bo69}:
\begin{equation}
    B(T_\lambda; J_2 \rightarrow J_1) \ = \ \frac{2J_1+1}{2J_2+1} \ B(T_\lambda; J_1 \rightarrow J_2)\ .
    \label{Bupdown}
\end{equation}
to obtain the value used herein.

We also present a set of 3pF parameters (listed under 'Here'), $R_c = 5.536$~fm, $a_c = 0.428$~fm and $w = 0.014$ which reproduce both the $R_{rms}$ of Ref.~\cite{Gi75} and the known $B(E2;2^+_1 \rightarrow 0^+_1)$. 
The parameter sets from the literature, chosen only to reproduce the $R_{rms}$, underestimate this by about 10\%. Thus, it is proven that a 3pF set can be found constrained by both pieces of data.

Ref.~\cite{Fe85} obtained a value $B(E3) = 0.131 \pm 0.015$~$e^2$b$^3$. This is the latest value as given in the current evaluation \cite{Ki02}. The value of $\beta_3 = 0.155$ used in Section~\ref{discuss}, chosen there to give the best fit to spectral data, reproduces this well when used with the 3pF parameters $R_c = 5.536$~fm, $a_c = 0.428$~fm and $w = 0.014$, validating that choice of $\beta_3$.

\section{Conclusions}
\label{conc}

The low-energy states of $^{115}$Cd have been investigated within the MCAS framework, describing the system as a neutron coupled to the low-lying states of $^{114}$Cd. The target nucleus was represented in terms of a vibrational model including quadrupole and octupole excitations. The first five states and first $3^-$ state of $^{114}$Cd were included in the coupled-channel calculation, allowing the effects of the different core excitations on the spectrum of 115Cd to be examined. 

The ground state of $^{115}$Cd, with $J^\pi = \sfrac{1}{2}^+$, is well reproduced within the MCAS calculation, with its description involving the ground and the quadrupole excitation of the $^{114}$Cd core.  A similar description applies to the first excited state, $J^\pi = \sfrac{11}{2}^-$, associated with the coupling of the neutron to the $0h\sfrac{11}{2}$ orbit. The inclusion of additional target states produces changes in the calculated spectrum, although their effects are not equally significant. In particular, the $4^+_1$ state of $^{114}$Cd has a very small effect on the calculated levels, whereas the inclusion of the $3^-_1$ state produces modest shifts in some of the predicted level clusters. The effects of the excited quadrupole configurations are also relevant to the description of the low-lying spectrum.

These results indicate that a vibrational description of the target provides a useful framework for understanding several low-lying states of $^{115}$Cd. However, the deviations of the excited states of $^{114}$Cd from a simple harmonic-vibrator scheme suggest that anharmonic effects or mixed-phonon configurations may need to be considered for a more complete description. The electromagnetic transition strengths discussed in this work provide additional constraints on the adopted description, although some calculated $B(E2)$ values differ substantially from the available experimental data.
The present results therefore support the usefulness of a vibrational coupled-channel representation for describing selected low-energy states of $^{115}$Cd, but do not establish that the $^{114}$Cd core behaves as a pure vibrator. Further investigations, including a more detailed analysis of the channel composition of the calculated states, would help clarify their structural interpretation.

\bibliography{MCAS_2026}
\end{document}